# Online Page Migration on Ring Networks in Uniform Model

**Amanj Khorramian*** and **Akira Matsubayashi†**

Graduate School of Natural Science and Technology
Kanazawa University, Kanazawa 920-1192, Japan

**Abstract** This paper explores the problem of page migration in ring networks. A ring network is a connected graph, in which each node is connected with exactly two other nodes. In this problem, one of the nodes in a given network holds a page of size $D$. This node is called the server and the page is a non-duplicable data in the network. Requests are issued by nodes to access the page one after another. Every time a new request is issued, the server must serve the request and may migrate to another node before the next request arrives. A service costs the distance between the server and the requesting node, and the migration costs the distance of the migration multiplied by $D$. The problem is to minimize the total costs of services and migrations. We study this problem in uniform model, for which the page has a unit size, i.e. $D = 1$. A 3.326-competitive algorithm improving the current best upper bound is designed. We show that this ratio is tight for our algorithm.

## 1 Introduction

Page migration (a.k.a. data migration, file migration) is a classic problem in the area of shared memory management on a network of processors having their own local memories. In this problem, a sequence of requests are issued by processors one by one to access the page that is a single shared data object. Every time a new request is issued, a processor holding the page, called a server, must serve the request through its communication with the requesting processor. After serving the request, the server may migrate to another processor before the next request is issued. A service and a migration are assumed to cost the distance of the service communication and the distance of the migration multiplied by the page size $D \geq 1$, respectively. The goal of the problem is to minimize the total costs of services and migrations. The page migration problem can also be viewed as the management of shared information among nodes of a distributed network (Bienkowski 2012). In this paper, we consider the page migration problem on ring networks, one of the most common network topologies.

The page migration problem was firstly studied by Black and Sleator, using the framework of online algorithms and competitive analysis (1989). The notions of online algorithms and competitive analysis are given in Section 2. The best published deterministic algorithm on general networks with any $D$ is 4.086-competitive and was proposed by Bartal, Charikar, and Indyk (2001), which is recently improved by a 4-competitive algorithm using a dynamic phase-based approach reported by Bienkowski, Byrka, and Mucha (arXiv:1609.00831v1). Better algorithms exist for restricted networks. Actually, 3-competitive deterministic algorithms were proposed for trees and uniform networks with any $D$ (Black et al. 1989), and for three points with $D \in \{1, 2\}$ (Chrobak et al. 1997; Matsubayashi 2015a). These 3-competitive algorithms are optimal because 3 is also a lower bound even on two points (Black et al. 1989). Further results on three points with $D \geq 3$ are a $(3 + 1/D)$-competitive deterministic algorithm and a lower bound of $3 + \Omega(1/D)$ (Matsubayashi 2015a). This algorithm implies a $(3 + 1/D)$-competitive deterministic algorithm on 3-node ring networks. However, we do not know any deterministic algorithm with competitiveness better than 4 even for the extremely simple topology of ring networks with more than three nodes. Therefore, we restrict also the page size $D$. As shown in the case of three points, the possible best competitiveness may depend on $D$. Actually, the upper bound of 4 for general networks with any $D$ can be reduced to $2 + \sqrt{2} \approx 3.414$ if $D = 1$ (Matsubayashi 2008). The setting of the unit page size $D = 1$ is often called the uniform model in the context of data management problems including the page migration problem (Bienkowski 2012; Maggs et al. 1997; Khorramian et al. 2016; Matsubayashi 2008; Meyer auf der Heide et al. 1999). In the uniform model, lower bounds of 3.1639 for general networks and 3.1213 for 5-node ring networks were known (Matsubayashi 2008). Although we did not know for a long time if a lower bound of $3 + \Omega(1)$ exists, where $\Omega$ notation is with respect to $D$, a lower bound of $3 + 7.4 \times 10^{-6}$ with any $D$ was recently proved (Matsubayashi 2015b).

---
* Corresponding author: khorramian@gmail.com,
Tel: +98-918-974-3123

† mbayashi@t.kanazawa-u.ac.jp

Our contribution in this paper is to propose a 3.326-competitive algorithm on ring networks with $D = 1$. I.e., we prove that the competitiveness of 3.414 for general networks in the uniform model can be improved to 3.326 on ring networks. Our algorithm inspects the distances among the current requesting node, the previous requesting node, and the current server node, then upon each request, either migrates the server to one of the two requesting nodes or keeps the server at its current location without migration. We define our algorithm and prove its competitiveness in Section 3 through a competitive analysis. We also prove the tightness of our analysis in Section 4.

Other previous results for the page migration problem are as follows. As for randomized algorithms against adaptive online adversaries, a 3-competitive algorithm for general networks was proposed in (Westbrook 1994) and the upper bound of 3 is also a lower bound on two points (Bartal et al. 1995). As for a randomized algorithm against oblivious adversaries, a $c(D)$-competitive algorithm for general networks was proposed in (Westbrook 1994), where $c(D)$ is a function that tends toward 2.618 as $D$ grows large. Moreover, $(2 + 1/2D)$-competitive algorithms for trees (Chrobak et al. 1997) and uniform networks (Lund et al. 1999) were known. These $(2 + 1/2D)$-competitive algorithms are optimal because $2 + 1/2D$ is also a lower bound for any algorithm against oblivious adversaries even on two points (Chrobak et al. 1997). The page migration problem is studied also on continuous metric spaces (Chrobak et al. 1997; Khorramian et al. 2016).

The rest of this paper is organized as follows. Some preliminaries and notations are given in Section 2. We define our algorithm on ring networks and prove its competitiveness of $\rho \approx 3.326$ in Section 3. In Section 4, we prove a lower bound of $\rho$ for our algorithm. We conclude the paper in Section 5. Because the exact value of $\rho$ is complicated, we present it in Appendix A.

## 2 Preliminaries

Let $G$ be a cycle graph and $d(a, b)$ denote the distance of a shortest path between $a$ and $b$. For a given initial page location $s_0 \in G$, a sequence of requests $r_1, \ldots, r_n \in G$, and a page size $D \in \mathbb{Z}^+$, the page migration problem is to compute page locations $s_1, \ldots, s_n \in G$ such that the cost function $\sum_{i=1}^{n} \bigl(\delta(s_{i-1}, r_i) + D \cdot \delta(s_{i-1}, s_i)\bigr)$ is minimized. We call $s_i$ the server before request $r_{i+1}$ occurs.

An online algorithm must compute $s_i$ without any information about the locations of $r_{i+1}, \ldots, r_n$. On the other hand, an offline algorithm may compute $s_i$ using the information about the entire sequence of requests $r_1, \ldots, r_n$. An adversary against an online algorithm $A$ generates a sequence of requests given to $A$, and computes an output sequence of server locations. If $A$ is deterministic, then the adversary generates requests using the definition of $A$, or equivalently, the information of the actual behavior of $A$, and computes its own output according to an optimal offline algorithm $OPT$. The deterministic algorithm $A$ is $c$-competitive if $cost_A(s_0, \sigma) \leq c \cdot cost_{OPT}(s_0, \sigma) + \alpha$ for the initial server $s_0$ and any sequence $\sigma$ of requests, where $cost_A$ and $cost_{OPT}$ are costs of $A$ and $OPT$, respectively, and $\alpha$ is a constant. For randomized online algorithms, there are two types of adversaries. An adversary is said to be oblivious if it generates requests in advance only using the definition of $A$, i.e., without any information about the random behavior of $A$ and computes its own output according to $OPT$. In contrast, an adaptive online adversary generates requests using information of the random behavior of $A$ and computes its own output in an online fashion. The competitiveness of a randomized online algorithm against oblivious or adaptive online adversaries is defined in a similar way to that of a deterministic online algorithm, except that expected values are used for randomized costs.

It is common to use a potential function $\Phi$ for proving the competitiveness of an online algorithm $A$. The potential function typically maps the situation at a point of time, such as the page locations of $A$ and $OPT$, to a real value. More specifically, we suitably divide the sequence of the online processes of $A$ and $OPT$ into certain events. Our goal is to define the value of $\Phi$ in such a way that the initial value of $\Phi$ is at most some constant $\mu$, $\Phi$ is always at least some constant $-\nu$, and that $\Delta cost_A + \Delta \Phi \leq c \cdot \Delta cost_{OPT}$ for any event, where $\Delta$ denotes the change of values by the event. Summing the inequality over all events for an initial server $s_0$ and a request sequence $\sigma$, we have $cost_A(s_0, \sigma) \leq c \cdot cost_{OPT}(s_0, \sigma) + \mu + \nu$, which means that $A$ is $c$-competitive.

## 3 Design and Analysis of Algorithm

For the page size $D = 1$, we propose a deterministic algorithm, called TriAct, on a ring network as defined in Figure 1. We set $r_0 = s_0$, the initial location of the server in the ring network $G$. For $i \geq 1$, upon the request $r_i$ at any node, there are three choices for the server to act according to the algorithm. The server keeps its current location at node $s_{i-1}$ or migrates to either $r_i$ or $r_{i-1}$. The decision is

```
L ← network length
x ← d(s_{i-1}, r_{i-1})
y ← d(s_{i-1}, r_i)
z ← d(r_{i-1}, r_i)
if z = x - y then s_i ← r_i                                                        {Case A}
else if z = y - x then s_i ← r_{i-1}                                               {Case B}
else if z = x + y then s_i ← s_{i-1}                                               {Case C}
else if y ≥ -(ρ-3)/(ρ-2) x + 1/2 L and y ≥ 2/ρ x + (ρ-2)/(2ρ) L then s_i ← r_{i-1} {Case D}
else if y ≤ (ρ-1)/2 x and y ≥ -ρ/(ρ-2) x + ρ/(2ρ-4) L then s_i ← r_i               {Case E}
else s_i ← s_{i-1}                                                                 {Case F}
server migrates from s_{i-1} to s_i
```

**Figure 1.** Definition of Algorithm TriAct upon the request $r_i$ to access the page at the server $s_{i-1}$ on a ring network

based on the distances among $s_{i-1}, r_i$, and $r_{i-1}$. There are six different cases to determine which action must be done.

In the ring topology, there are exactly two paths between each pair of nodes $a, b$. Let $\delta(a, b)$ denote a shortest path between $a, b$. The length of $\delta(a, b)$ equals to the distance $d(a, b)$. Let $L$ denote the length of the ring, then it is obvious that $0 \leq d(a,b) \leq L/2$. In our calculations, we set $x = d(s_{i-1}, r_{i-1})$, $y = d(s_{i-1}, r_i)$, and $z = d(r_{i-1}, r_i)$.

If $\delta(s_{i-1}, r_{i-1})$ and $\delta(s_{i-1}, r_i)$ share any edge of the network, then either Case A or Case B follows by the algorithm. If $\delta(r_{i-1}, r_i) = \delta(s_{i-1}, r_{i-1}) \cup \delta(s_{i-1}, r_i)$ then Case C follows. Figure 2 shows an example for each of these three cases.

For the rest of cases, we have $L = d(r_{i-1}, r_i) + d(s_{i-1}, r_{i-1}) + d(s_{i-1}, r_i) = x + y + z$, and the algorithm separates all possible conditions among distances into three Cases D, E, and F, to decide the action of server for migration. Since $L$ is a constant value, we calculate $z$ as a function of $x$ and $y$. The conditions of Cases D, E, and F are shown in Figure 3.

We show that TriAct is $\rho$-competitive with $\rho \approx 3.326$ in Theorem 1 below. The exact value of $\rho$ is provided in Appendix A. We use a potential function $\Phi$ to prove the theorem. We separate the online events into two parts to show that $\Delta cost_{TriAct} + \Delta \Phi - 3.326 \Delta cost_{OPT} \leq 0$ follows in every case. The proof for Cases A-E are straightforward. Our analysis for Case F uses a different technique, because we need to consider two consecutive requests in that case to complete the proof.

**Theorem 1.** *TriAct is $\rho$-competitive for $D = 1$, where $\rho \approx 3.326$ is the positive solution of $-\rho^4 + 4\rho^3 + \rho^2 - 18\rho + 24 = 0$. (See Appendix A for the exact value of $\rho$.)*

**Proof.** We use the potential function $\Phi$, for $OPT$'s server locations $t_1, \ldots, t_k$, TriAct's server locations $s_1, \ldots, s_k$, and request locations $r_1, \ldots, r_k$. We define

$$\Phi(s_i, r_i, t_i) = \frac{\rho}{2} \cdot \big(d(s_i, t_i) + d(r_i, t_i)\big) + \left(\frac{\rho}{2} - 1\right) d(s_i, r_i)$$

We separately consider the events in two parts. The first includes the migration costs incurred by $OPT$, and the second covers the service costs incurred by $OPT$ together with the migration and service costs incurred by TriAct. Let

$$\Delta_1 = \Phi(s_i, r_i, t_i) - \Phi(s_i, r_i, t_{i-1}) - \rho \cdot d(t_{i-1}, t_i),$$

and

$$\Delta_2 = d(s_{i-1}, r_i) + d(s_{i-1}, s_i) + \Phi(s_i, r_i, t_{i-1}) - \Phi(s_{i-1}, r_{i-1}, t_{i-1}) - \rho \cdot d(t_{i-1}, r_i).$$

In order to prove that $\Delta cost_{PQ} + \Delta \Phi - \rho \cdot \Delta cost_{OPT} \leq 0$ follows in both parts, it suffices to show that $\Delta_1 \leq 0$ and $\Delta_2 \leq 0$.

**Analysis of part 1:**

For the first part,
$$\Delta_1 = \frac{\rho}{2} \cdot \big(d(s_i, t_i) - d(s_i, t_{i-1}) + d(r_i, t_i) - d(r_i, t_{i-1})\big) - \rho \cdot d(t_{i-1}, t_i).$$

By triangle inequality, we have
$$\Delta_1 \leq \frac{\rho}{2} \cdot \big(d(t_{i-1}, t_i) + d(t_{i-1}, t_i)\big) - \rho \cdot d(t_{i-1}, t_i) = 0.$$

**Analysis of part 2:**

For the second part, we have
$$\begin{aligned}
\Delta_2 = &\ d(s_{i-1}, r_i) + d(s_{i-1}, s_i) + \frac{\rho}{2} d(s_i, t_{i-1}) \\
&+ \frac{\rho}{2} d(r_i, t_{i-1}) + \left(\frac{\rho}{2} - 1\right) d(s_i, r_i) \\
&- \frac{\rho}{2} d(s_{i-1}, t_{i-1}) - \frac{\rho}{2} d(r_{i-1}, t_{i-1}) \\
&- \left(\frac{\rho}{2} - 1\right) d(s_{i-1}, r_{i-1}) - \rho \cdot d(t_{i-1}, r_i) \\
= &\ d(s_{i-1}, r_i) + d(s_{i-1}, s_i) + \frac{\rho}{2} d(s_i, t_{i-1}) \\
&- \frac{\rho}{2} d(r_i, t_{i-1}) + \left(\frac{\rho}{2} - 1\right) d(s_i, r_i) \\
&- \frac{\rho}{2} d(s_{i-1}, t_{i-1}) - \frac{\rho}{2} d(r_{i-1}, t_{i-1}) \\
&- \left(\frac{\rho}{2} - 1\right) d(s_{i-1}, r_{i-1}).
\end{aligned} \qquad (1)$$

TriAct has three choices of $s_i$, i.e., $r_i$, or $r_{i-1}$, or $s_{i-1}$. For these choices, we separately derive upper bounds of $\Delta_2$.

**Upper bound of $\Delta_2$ for the action $s_i \leftarrow r_i$:**

For the action of migrating the server to the current request location, it follows from (1) that

$$\begin{aligned}\Delta_2 &= d(s_{i-1}, r_i) + d(s_{i-1}, r_i) + \frac{\rho}{2} d(r_i, t_{i-1}) \\ &\quad - \frac{\rho}{2} d(r_i, t_{i-1}) + \left(\frac{\rho}{2} - 1\right) d(r_i, r_i) \\ &\quad - \frac{\rho}{2} d(s_{i-1}, t_{i-1}) - \frac{\rho}{2} d(r_{i-1}, t_{i-1}) \\ &\quad - \left(\frac{\rho}{2} - 1\right) d(s_{i-1}, r_{i-1}) \\ &= 2d(s_{i-1}, r_i) - \frac{\rho}{2} d(s_{i-1}, t_{i-1}) \\ &\quad - \frac{\rho}{2} d(r_{i-1}, t_{i-1}) - \left(\frac{\rho}{2} - 1\right) d(s_{i-1}, r_{i-1}) \\ &\leq 2d(s_{i-1}, r_i) - \frac{\rho}{2} d(s_{i-1}, r_{i-1}) \\ &\quad - \left(\frac{\rho}{2} - 1\right) d(s_{i-1}, r_{i-1}) \\ &= (1-\rho) d(s_{i-1}, r_{i-1}) + 2d(s_{i-1}, r_i) \\ &= (1-\rho) x + 2y. \end{aligned} \quad (2)$$

Here, we used the triangle inequality $d(s_{i-1}, t_{i-1}) + d(r_{i-1}, t_{i-1}) \geq d(s_{i-1}, r_{i-1})$. We note that this upper bound of $\Delta_2$ is used for Cases A and E.

**Upper bound of $\Delta_2$ for the action $s_i \leftarrow r_{i-1}$:**

For the action of migrating the server to the previous request location, it follows from (1) that

$$\begin{aligned}\Delta_2 &= d(s_{i-1}, r_i) + d(s_{i-1}, r_{i-1}) \\ &\quad + \frac{\rho}{2} d(r_{i-1}, t_{i-1}) - \frac{\rho}{2} d(r_i, t_{i-1}) \\ &\quad + \left(\frac{\rho}{2} - 1\right) d(r_{i-1}, r_i) - \frac{\rho}{2} d(s_{i-1}, t_{i-1}) \\ &\quad - \frac{\rho}{2} d(r_{i-1}, t_{i-1}) - \left(\frac{\rho}{2} - 1\right) d(s_{i-1}, r_{i-1}) \\ &= d(s_{i-1}, r_i) + \left(2 - \frac{\rho}{2}\right) d(s_{i-1}, r_{i-1}) \\ &\quad - \frac{\rho}{2} d(t_{i-1}, r_i) + \left(\frac{\rho}{2} - 1\right) d(r_{i-1}, r_i) \\ &\quad - \frac{\rho}{2} d(s_{i-1}, t_{i-1}) \\ &\leq \left(2 - \frac{\rho}{2}\right) d(s_{i-1}, r_{i-1}) \\ &\quad + \left(1 - \frac{\rho}{2}\right) d(s_{i-1}, r_i) \\ &\quad + \left(\frac{\rho}{2} - 1\right) d(r_{i-1}, r_i) \\ &= \left(2 - \frac{\rho}{2}\right) x + \left(1 - \frac{\rho}{2}\right) y + \left(\frac{\rho}{2} - 1\right) z. \end{aligned} \quad (3)$$

Here, we used the triangle inequality $d(t_{i-1}, r_i) + d(s_{i-1}, t_{i-1}) \geq d(s_{i-1}, r_i)$. We note that this upper bound of $\Delta_2$ is used for Cases B and D.

**Upper bound of $\Delta_2$ for the action $s_i \leftarrow s_{i-1}$:**

For the action of no migration, it follows from (1) that

$$\begin{aligned}\Delta_2 &= d(s_{i-1}, r_i) + d(s_{i-1}, s_{i-1}) \\ &\quad + \frac{\rho}{2} d(s_{i-1}, t_{i-1}) - \frac{\rho}{2} d(r_i, t_{i-1}) \\ &\quad + \left(\frac{\rho}{2} - 1\right) d(s_{i-1}, r_i) - \frac{\rho}{2} d(s_{i-1}, t_{i-1}) \\ &\quad - \frac{\rho}{2} d(r_{i-1}, t_{i-1}) - \left(\frac{\rho}{2} - 1\right) d(s_{i-1}, r_{i-1}) \\ &= \frac{\rho}{2} d(s_{i-1}, r_i) - \frac{\rho}{2} d(r_i, t_{i-1}) \\ &\quad - \frac{\rho}{2} d(r_{i-1}, t_{i-1}) - \left(\frac{\rho}{2} - 1\right) d(s_{i-1}, r_{i-1}) \\ &\leq \frac{\rho}{2} d(s_{i-1}, r_i) - \frac{\rho}{2} d(r_{i-1}, r_i) \\ &\quad - \left(\frac{\rho}{2} - 1\right) d(s_{i-1}, r_{i-1}) \\ &= \left(1 - \frac{\rho}{2}\right) x + \frac{\rho}{2} y - \frac{\rho}{2} z. \end{aligned} \quad (4)$$

Here, we used the triangle inequality $d(r_i, t_{i-1}) + d(r_{i-1}, t_{i-1}) \geq d(r_{i-1}, r_i)$. We note that this upper bound of $\Delta_2$ is used for Cases C and F.

**Analysis for Cases A, B, and C:**

In Case A, since $y = x - z \leq x$ and $\rho > 3$, it follows from (2) that

$$\Delta_2 \leq (1-\rho) x + 2y \leq (3-\rho) x < 0. \quad (5)$$

In Case B, since $z = y - x$ and $\rho > 3$, it follows from (3) that

$$\begin{aligned}\Delta_2 &\leq \left(2 - \frac{\rho}{2}\right) x + \left(1 - \frac{\rho}{2}\right) y + \left(\frac{\rho}{2} - 1\right) z \\ &= \left(2 - \frac{\rho}{2}\right) x + \left(1 - \frac{\rho}{2}\right) y + \left(\frac{\rho}{2} - 1\right) (y - x) \\ &= (3-\rho) x < 0. \end{aligned} \quad (6)$$

In Case C, since $z = x + y$ and $\rho > 1$, it follows from (4) that

$$\begin{aligned}\Delta_2 &\leq \left(1 - \frac{\rho}{2}\right) x + \frac{\rho}{2} y - \frac{\rho}{2} z \\ &= \left(1 - \frac{\rho}{2}\right) x + \frac{\rho}{2} y - \frac{\rho}{2} (x+y) \\ &= (1-\rho) x < 0. \end{aligned} \quad (7)$$

Therefore, Theorem 1 holds in Cases A, B, and C.

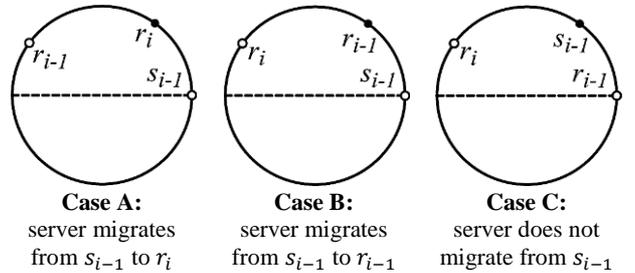

Case A: server migrates from $s_{i-1}$ to $r_i$

Case B: server migrates from $s_{i-1}$ to $r_{i-1}$

Case C: server does not migrate from $s_{i-1}$

**Figure 2.** Examples of Cases A, B, and C. Filled circles represent destinations of the server, i.e., $s_i$, upon the request $r_i$

**Analysis for Cases D and E:**

For Cases D and E, we have $z = L - x - y$. The conditions of these cases are defined using four functions $y_1$-$y_4$ of $x$, where

$$y_1 = -\frac{\rho - 3}{\rho - 2}x + \frac{L}{2}, \tag{8}$$

$$y_2 = \frac{2}{\rho}x + \frac{\rho - 2}{\rho} \cdot \frac{L}{2}, \tag{9}$$

$$y_3 = \frac{\rho - 1}{2}x, \text{ and} \tag{10}$$

$$y_4 = \frac{\rho}{\rho - 2}\left(\frac{L}{2} - x\right). \tag{11}$$

In Figure 3, the separate regions represent the conditions of Cases D, E, and F.

In Case D, since $y \geq y_1$, it follows from (3) and (8) that

$$\Delta_2 \leq \left(2 - \frac{\rho}{2}\right)x + \left(1 - \frac{\rho}{2}\right)y$$
$$+ \left(\frac{\rho}{2} - 1\right)(L - x - y)$$
$$= (3 - \rho)x + (2 - \rho)y + \left(\frac{\rho}{2} - 1\right)L \tag{12}$$
$$\leq (3 - \rho)x + (2 - \rho)\left(-\frac{\rho - 3}{\rho - 2}x + \frac{L}{2}\right)$$
$$+ \left(\frac{\rho}{2} - 1\right)L$$
$$= 0.$$

In Case E, since $y \leq y_3$, it follows from (2) and (10) that

$$\Delta_2 \leq (1 - \rho)x + 2y \tag{13}$$
$$\leq (1 - \rho)x + 2\left(\frac{\rho - 1}{2}x\right) = 0.$$

Therefore, Theorem 1 holds in Cases D and E.

**Analysis for Case F:**

For Case F, we also have $z = L - x - y$ and the conditions of Case F as shown in Figure 3. It follows from (4) that

$$\Delta_2 \leq \left(1 - \frac{\rho}{2}\right)x + \frac{\rho}{2}y - \frac{\rho}{2}z$$
$$= \left(1 - \frac{\rho}{2}\right)x + \frac{\rho}{2}y$$
$$- \frac{\rho}{2}(L - x - y) = \rho y + x - \frac{\rho}{2}L. \tag{14}$$

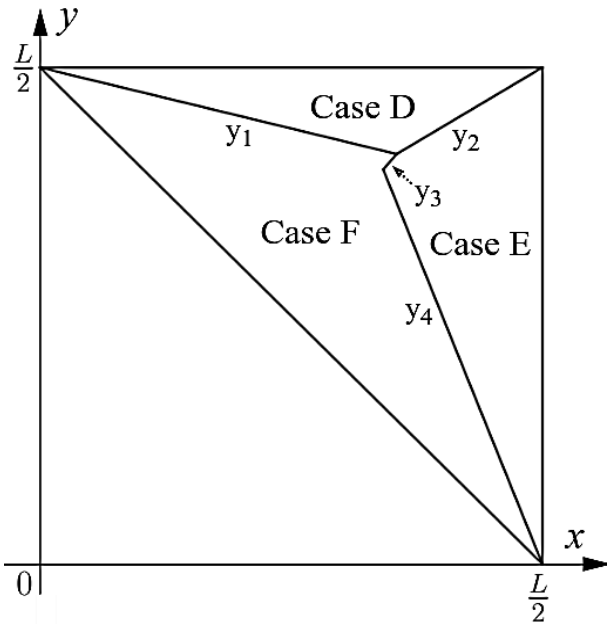

**Figure 3.** Conditions of Cases D, E, and F

This is at most 0 if $y \leq y_5$, where

$$y_5 = \frac{L}{2} - \frac{x}{\rho}. \tag{15}$$

Therefore, Theorem 1 holds if $y \leq y_5$ in Case F. However, if $y > y_5$, i.e., if $x$ and $y$ are in the grey region in Figure 4, then $\Delta_2 > 0$. Instead of bounding $\Delta_2$, for the case $y > y_5$, we bound the sum of $\Delta_2$ and $\Delta_2'$, which is defined as the value of $\Delta_2$ for the next request $r_{i+1}$. Specifically,

$$\Delta_2' = d(s_i, r_{i+1}) + d(s_i, s_{i+1}) + \Phi(s_{i+1}, r_{i+1}, t_i)$$
$$- \Phi(s_i, r_i, t_i) - \rho \cdot d(t_i, r_{i+1}).$$

It should be noted that Theorem 1 holds if $\Delta_2 + \Delta_2' < 0$ for all six cases of $r_{i+1}$, and for all x and y in the grey region in Figure 4. We also note that

$$s_i = s_{i-1} \tag{16}$$

in Case F for $r_i$. In the rest of the proof, we show $\Delta_2 + \Delta_2' \leq 0$ for all six cases of $r_{i+1}$ and for all $x$ and $y$ with $\Delta_2 > 0$.

In Cases A, B, and C for $r_{i+1}$, if follows from (5), (6), (7) and (16) that $\Delta_2' \leq (3 - \rho)d(s_i, r_i) = (3 - \rho)d(s_{i-1}, r_i) = (3 - \rho)y$. Therefore, since $y \leq y_1$, it follows from (14) and (8) that

$$\Delta_2 + \Delta_2' \leq \rho y + x - \frac{\rho}{2}L - (\rho - 3)y$$
$$= 3y + x - \frac{\rho}{2}L$$
$$\leq 3\left(-\frac{\rho - 3}{\rho - 2}x + \frac{L}{2}\right) + x - \frac{\rho}{2}L$$
$$= \frac{7 - 2\rho}{\rho - 2}x - (\rho - 3)\frac{L}{2}$$
$$\leq \frac{7 - 2\rho}{\rho - 2} \cdot \frac{L}{2} - (\rho - 3)\frac{L}{2}$$
$$= -\frac{\rho^2 - 3\rho - 1}{\rho - 2} \cdot \frac{L}{2},$$

which is negative since $\rho > \frac{3 + \sqrt{13}}{2} \approx 3.303$.

In Case D for $r_{i+1}$, it follows from (12) and (16) that

$$\Delta_2' \leq -(\rho - 2)d(s_i, r_{i+1}) - (\rho - 3)d(s_i, r_i)$$
$$+ \left(\frac{\rho}{2} - 1\right)L$$
$$= -(\rho - 2)d(s_i, r_{i+1}) - (\rho - 3)d(s_{i-1}, r_i)$$
$$+ \left(\frac{\rho}{2} - 1\right)L$$
$$= -(\rho - 2)d(s_i, r_{i+1}) - (\rho - 3)y + \left(\frac{\rho}{2} - 1\right)L. \tag{17}$$

Moreover, it follows from (9) and (16) that

$$d(s_i, r_{i+1}) \geq \frac{2}{\rho}d(s_i, r_i) + \frac{\rho - 2}{\rho} \cdot \frac{L}{2}$$
$$= \frac{2}{\rho}d(s_{i-1}, r_i) + \frac{\rho - 2}{\rho} \cdot \frac{L}{2}$$
$$= \frac{2}{\rho}y + \frac{\rho - 2}{\rho} \cdot \frac{L}{2}. \tag{18}$$

Therefore, it follows from (14), (17), and (18) that

$$\Delta_2 + \Delta_2' \le \rho y + x - \frac{\rho}{2}L - (\rho - 2)d(s_i, r_{i+1})$$
$$- (\rho - 3)y + \left(\frac{\rho}{2} - 1\right)L$$
$$\le 3y + x - L - (\rho - 2)\left(\frac{2}{\rho}y + \frac{\rho - 2}{\rho} \cdot \frac{L}{2}\right)$$
$$= \frac{\rho + 4}{\rho}y + x - \frac{\rho^2 - 2\rho + 4}{\rho} \cdot \frac{L}{2}. \quad (19)$$

The value $C = \frac{\rho+4}{\rho}y + x$ is maximized at $p$, at which $y_1$ and $y_3$ intersect in Figure 4. This is because for the function $y = -\frac{\rho}{\rho+4}x + \frac{\rho}{\rho+4}C$, its slope $-\frac{\rho}{\rho+4}$ is negative and less than the slope $-\frac{\rho-3}{\rho-2}$ of $y_1$ for $\rho \approx 3.326$. Therefore, we have

$$\Delta_2 + \Delta_2' \le \frac{\rho + 4}{\rho} \cdot \frac{\rho^2 - 3\rho + 2}{\rho^2 - \rho - 4} \cdot \frac{L}{2} + \frac{\rho - 2}{\rho^2 - \rho - 4}L$$
$$- \frac{\rho^2 - 2\rho + 4}{\rho} \cdot \frac{L}{2}$$
$$= \frac{-\rho^4 + 4\rho^3 + \rho^2 - 18\rho + 24}{\rho(\rho^2 - \rho - 4)} \cdot \frac{L}{2},$$

which is equal to 0 by the assumption of $\rho$ in Theorem 1.

In Case E for $r_{i+1}$, it follows from (13) and (16) that
$$\Delta_2' \le 2d(s_i, r_{i+1}) - (\rho - 1)d(s_i, r_i)$$
$$= 2d(s_i, r_{i+1}) - (\rho - 1)d(s_{i-1}, r_i)$$
$$= 2d(s_i, r_{i+1}) - (\rho - 1)y. \quad (20)$$

Moreover, it follows from (9) and (16) that
$$d(s_i, r_{i+1}) \le \frac{2}{\rho}d(s_i, r_i) + \frac{\rho - 2}{\rho} \cdot \frac{L}{2}$$
$$= \frac{2}{\rho}d(s_{i-1}, r_i) + \frac{\rho - 2}{\rho} \cdot \frac{L}{2}$$
$$\le \frac{2}{\rho}y + \frac{\rho - 2}{\rho} \cdot \frac{L}{2}. \quad (21)$$

Therefore, it follows from (14), (20), and (21) that
$$\Delta_2 + \Delta_2' \le \rho y + x - \frac{\rho}{2}L + 2d(s_i, r_{i+1}) - (\rho - 1)y$$

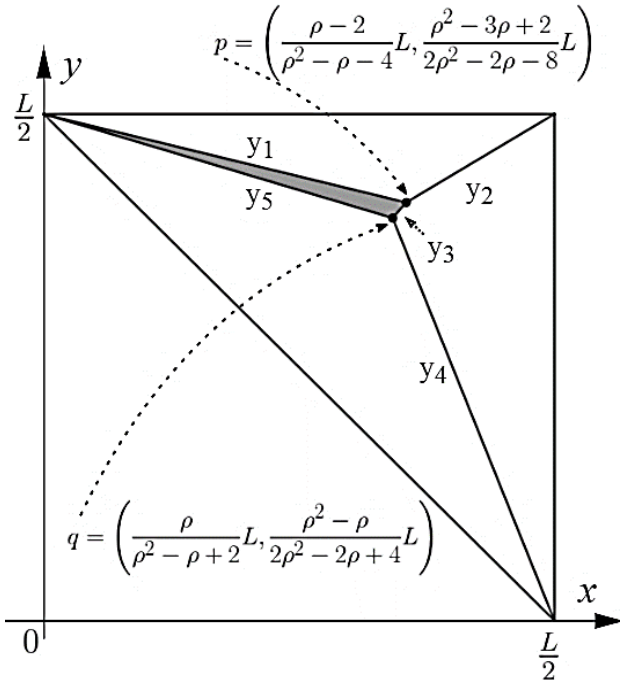

**Figure 4.** Representation of positive region of $\Delta_2$ in Case F

$$\le y + x - \frac{\rho}{2}L + 2\left(\frac{2}{\rho}y + \frac{\rho - 2}{\rho} \cdot \frac{L}{2}\right)$$
$$= \frac{\rho + 4}{\rho}y + x - \frac{\rho^2 - 2\rho + 4}{\rho} \cdot \frac{L}{2},$$

which is identical with (19). Hence, we can obtain $\Delta_2 + \Delta_2' \le 0$ as done for (19).

In Case F for $r_{i+1}$, it follows from (14) and (16) that
$$\Delta_2' \le \rho d(s_i, r_{i+1}) + d(s_i, r_i) - \frac{\rho}{2}L$$
$$= \rho d(s_i, r_{i+1}) + d(s_{i-1}, r_i) - \frac{\rho}{2}L$$
$$= \rho d(s_i, r_{i+1}) + y - \frac{\rho}{2}L. \quad (22)$$

For $x$ and $y$ in the grey region in Figure 4, it follows that $y \ge \frac{\rho^2 - \rho}{\rho^2 - \rho + 2} \cdot \frac{L}{2}$, which is the $y$-coordinate of $q$ at which $y_5$ and $y_3$ intersect. This implies that $d(s_i, r_i)$ is larger than the $x$-coordinate $\frac{\rho - 2}{\rho^2 - \rho - 4}L$ of $p$. We can verify this by $d(s_i, r_i) = d(s_{i-1}, r_i) = y$ and

$$\frac{\rho^2 - \rho}{\rho^2 - \rho + 2} \cdot \frac{L}{2} - \frac{\rho - 2}{\rho^2 - \rho - 4}L$$
$$= \frac{\rho^4 - 4\rho^3 + 3\rho^2 - 4\rho + 8}{(\rho - 2)(\rho + 1)(\rho^2 - \rho - 4)} \cdot \frac{L}{2}$$
$$= \frac{(\rho^4 - 4\rho^3 - \rho^2 + 18\rho - 24) + (4\rho^2 - 22\rho + 32)}{(\rho - 2)(\rho + 1)(\rho^2 - \rho - 4)} \cdot \frac{L}{2}$$
$$= \frac{4\rho^2 - 22\rho + 32}{(\rho - 2)(\rho + 1)(\rho^2 - \rho - 4)} \cdot \frac{L}{2},$$

which is positive for $\rho > \frac{1+\sqrt{17}}{2} \approx 2.56$. Here, we have used the assumption $\rho^4 - 4\rho^3 - \rho^2 + 18\rho - 24 = 0$ in Theorem 1. Therefore, it follows from (11) that

$$d(s_i, r_{i+1}) \le \frac{\rho}{\rho - 2}\left(\frac{L}{2} - d(s_i, r_i)\right)$$
$$= \frac{\rho}{\rho - 2}\left(\frac{L}{2} - d(s_{i-1}, r_i)\right)$$
$$= \frac{\rho}{\rho - 2}\left(\frac{L}{2} - y\right). \quad (23)$$

Therefore, it follows from (14), (22), and (23) that
$$\Delta_2 + \Delta_2' \le \rho y + x - \frac{\rho}{2}L + \rho d(s_i, r_{i+1}) + y - \frac{\rho}{2}L$$
$$\le (\rho + 1)y + x - \rho L + \rho \cdot \frac{\rho}{\rho - 2}\left(\frac{L}{2} - y\right)$$
$$= -\frac{\rho + 2}{\rho - 2}y + x - \frac{\rho(\rho - 4)}{\rho - 2} \cdot \frac{L}{2}.$$

The value $C = -\frac{\rho+2}{\rho-2}y + x$ is maximized at $q$. This is because for the function $y = \frac{\rho-2}{\rho+2}x - \frac{\rho-2}{\rho+2}C$, its slope $\frac{\rho-2}{\rho+2}$ is positive and less than the slope $\frac{\rho-1}{2}$ of $y_3$. Therefore, we have

$$\Delta_2 + \Delta_2' \le -\frac{\rho + 2}{\rho - 2} \cdot \frac{\rho(\rho - 1)}{\rho^2 - \rho + 2} \cdot \frac{L}{2} + \frac{\rho}{\rho^2 - \rho + 2}L$$
$$- \frac{\rho(\rho - 4)}{\rho - 2} \cdot \frac{L}{2}$$
$$= -\frac{\rho(\rho - 3)}{\rho - 2} \cdot \frac{L}{2} < 0.$$

Thus, the proof of Theorem 1 is completed. ∎

## 4 Tightness of Analysis

In Section 3, we prove that TriAct is $\rho$-competitive with $\rho \approx 3.326$. In this section, we show that the lower bound of the algorithm is also $\rho$. This means that the exact competitive ratio of the algorithm is $\rho$. We introduce an adversary through Theorem 2 and prove the existence of such a lower bound. The adversary makes a special sequence of requests on four nodes on a ring network against TriAct, such that upon any request either the condition of Case B or the condition of Case E holds.

**Theorem 2.** *For a sufficiently large integer $n$, there exists a request sequence $\sigma = r_1, \ldots, r_{4n}$ and four request nodes $s$, $a$, $b$, and $c$ on a ring network, such that $cost_{PQ}(s, \sigma) \geq \rho \cdot cost_{OPT}(s, \sigma)$ for $\rho \approx 3.326$ and the initial server node is $s$.*

**Proof.** The adversary sets $d(s,b) = \frac{\rho^2 - 3\rho + 2}{2\rho^2 - 2\rho - 8}L$ and $d(s,a) = d(b,c) = \frac{\rho - 2}{\rho^2 - \rho - 4}L$, such that neither $\delta(s,a)$ nor $\delta(b,c)$ has an edge of $\delta(s,b)$. Moreover, request $r_{4k+1}$ is at $a$, request $r_{4k+2}$ is at $b$, request $r_{4k+3}$ is at $c$, and request $r_{4k+4}$ is at $s$, where $0 \leq k < n$ (see Figure 5).

We first calculate the total cost incurred by TriAct, $cost_{TriAct}(s, \sigma)$. The request $r_{4k+1}$ is served with the cost of $d(s,a)$ and the server does not move from $s$, because Case B holds. The request $r_{4k+2}$ is served with the cost of $d(s,b)$ and the server migrates from $s$ to $b$ with the cost of $d(s,b)$, because Case E holds. The request $r_{4k+3}$ is served with the cost of $d(b,c)$ and the server does not move from $b$, because Case B holds. The request $r_{4k+4}$ is served with the cost of $d(b,s)$ and the server migrates from $b$ to $s$ with a cost of $d(b,s)$, because Case E holds. Therefore, we have
$$cost_{TriAct}(s, \sigma)$$
$$= n\big(d(s,a) + 2d(s,b) + d(b,c) + 2d(b,s)\big)$$
$$= n\big(2d(s,a) + 4d(s,b)\big).$$
Now we calculate the total cost incurred by a specific algorithm $ALG$, $cost_{ALG}(s, \sigma)$ that is at least $cost_{OPT}(s, \sigma)$. $ALG$ migrates the server from $s$ to $a$ before any request occurs, with the cost of $d(s,a)$. The request $r_{4k+1}$ is served with no cost and the server does not migrate from $a$. The request $r_{4k+2}$ is served with the cost of $d(a,b)$ and the server migrates from $a$ to $c$ with a cost of $d(a,c)$. The request $r_{4k+3}$ is served with no cost and the server does not migrate from $c$. The request $r_{4k+4}$ is served with the cost of $d(c,s)$ and the server migrates from $c$ to $a$ with a cost of $d(c,a)$. Therefore, we have
$$cost_{ALG}(s, \sigma)$$
$$= d(s,a) + n\big(d(a,b) + d(a,c) + d(c,s) + d(c,a)\big)$$
$$= d(s,a) + n\big(d(b,c) + d(s,a)\big)$$
$$= d(s,a) + n\big(2d(s,a)\big).$$
Since
$$\lim_{n \to \infty} \frac{n\big(2d(s,a) + 4d(s,b)\big)}{d(s,a) + n\big(2d(s,a)\big)}$$
$$= \frac{d(s,a) + 2d(s,b)}{d(s,a)}$$
$$= \frac{\frac{\rho-2}{\rho^2-\rho-4}L + 2 \cdot \frac{\rho^2-3\rho+2}{2\rho^2-2\rho-8}L}{\frac{\rho-2}{\rho^2-\rho-4}L} = \rho,$$
then $\frac{cost_{TriAct}(s,\sigma)}{cost_{OPT}(s,\sigma)}$ is at least $\rho$. This completes the proof. ∎

## 5 Concluding Remarks

In this paper, we proposed a deterministic algorithm for the page migration problem in ring networks with unit page size. The competitive ratio $\rho \approx 3.326$ of the algorithm is an improvement on the previous competitiveness of 3.414 in our setting. A tight example was found to express a lower bound of $\rho$ for the algorithm. If possible, one could seek to find an algorithm to cover a larger $D$ with better competitiveness than 4.

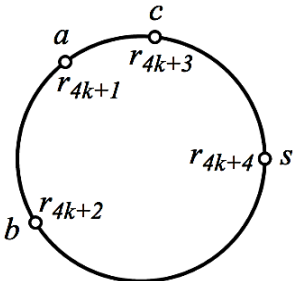

**Figure 5.** Representation of adversary model against TriAct

**Appendix A**

The exact value of $\rho$ is the positive solution of the equation $-\rho^4 + 4\rho^3 + \rho^2 - 18\rho + 24 = 0$. The strict solution is

$$\rho = 1 - \frac{\sqrt{9\sqrt[3]{\lambda^{\frac{3}{2}}} + 42\sqrt[3]{\lambda} - 71}}{6\sqrt[6]{\lambda}} + \frac{\sqrt{-\sqrt[3]{\lambda} + \frac{48\sqrt[6]{\lambda}}{\sqrt{9\sqrt[3]{\lambda^{\frac{3}{2}}} + 42\sqrt[3]{\lambda} - 71}} + \frac{71}{9\sqrt[3]{\lambda}} + \frac{28}{3}}}{2},$$

where $\lambda = \frac{2\sqrt{13438}}{3} - \frac{1999}{27}$.